\documentclass{article}

\usepackage{arxiv}

\usepackage[utf8]{inputenc} 
\usepackage[T1]{fontenc}    
\usepackage{url}            
\usepackage{booktabs}       
\usepackage{amsfonts}       
\usepackage{nicefrac}       
\usepackage{microtype}      
\usepackage{lipsum}
\usepackage{graphicx}
\usepackage{natbib}
\usepackage{amsthm}  
\usepackage{amsmath}
\usepackage{amssymb}
\usepackage{array}
\usepackage{boldline}
\usepackage{float}

\newtheorem{definition}{Definition}

\graphicspath{ {./images/} }

\title{A concordance coefficient for lattice data: \\An application to poverty indices in Chile}

\author{
  Ronny Vallejos \\
  Department of Mathematics\\
  Universidad T\'ecnica Federico Santa Maria\\
  Valpara\'iso, Chile \\
  \texttt{ronny.vallejos@usm.cl} \\
   \And
 Clemente Ferrer \\
  Department of Mathematics\\
  Universidad T\'ecnica Federico Santa Maria\\
  Valpara\'iso, Chile \\
  \texttt{clemente.ferrer@usm.cl} \\
  \And
 Jorge Mateu \\
  Department of Mathematics\\
  University Jaume I\\
  Castellón, Spain \\
  \texttt{mateu@uji.es} \\
}

\begin{document}
\maketitle
\begin{abstract}
This paper introduces a novel coefficient for measuring agreement between two lattice sequences observed in the same areal units, motivated by the analysis of different methodologies for measuring poverty rates in Chile. Building on the multivariate concordance coefficient framework, our approach accounts for dependencies in the multivariate lattice process using a non-negative definite matrix of weights, assuming a Multivariate Conditionally Autoregressive (GMCAR) process. We adopt a Bayesian perspective for inference, using summaries from Bayesian estimates. The methodology is illustrated through an analysis of poverty rates in the Metropolitan and Valparaíso regions of Chile, with High Posterior Density (HPD) intervals provided for the poverty rates. This work addresses a methodological gap in the understanding of agreement coefficients and enhances the usability of these measures in the context of social variables typically assessed in areal units.
\end{abstract}


\section{Introduction}
Concordance coefficients, or agreement measures, have gained increasing relevance over the past two decades due to their importance in validating consistency between two or more methods.
Specifically, when a new assay or instrument is developed, it is important to assess whether the new assay can reproduce the results of a traditional gold-standard method \citep{West:1973, Bauer:1981}. It is well established in the literature that classical techniques, such as the Pearson correlation coefficient, paired t-test, least squares analysis of slope and intercept, coefficient of variation, and intraclass correlation coefficient, do not offer new insights in this context. This is because the discrepancy between methods is measured relative to the $45^{\circ}$ line through the origin. Building on this principle, \cite{Lin:1989} developed the concordance correlation coefficient to quantify the discrepancy between two sets of measurements generated by different mechanisms.

Several extensions and generalizations of Lin's result have been studied. For instance, \cite{King:2007} introduced a new repeated measures concordance coefficient, which not only possesses the necessary properties to measure the overall agreement between two $n\times 1$ vectors of random variables but also offers a more intuitive appeal compared to the methods previously studied by \cite{Chinchilli:1996}. Later, \cite{Hiriote:2011} extended Lin's concordance coefficient to multivariate observations in the context of repeated measures.
 \cite{Leal:2019} studied  Lin's type coefficient in the context of local influence assessment. Additionally, \cite{Leal:2019} explored a Lin-type coefficient in the assessment of local influence. A  compendium of various techniques used in statistical agreement analysis can be found in \cite{Lin:2012} and \cite{Chou:2017}

Another approach to evaluating agreement between two sets of measurements is the probability of agreement (PA), first introduced in a series of papers by \cite{Stevens:2017}, \cite{Stevens:2018}, and \cite{Stevens:2020}. \cite{deCastro:2021} developed Bayesian PA methods to compare measurement systems with either homoscedastic or heteroscedastic errors. It is worth noting that the literature on PA is extensive, with numerous applications across various scenarios having been published.

Extensions of agreement coefficients for the analysis of spatial data have been the focus of recent investigations. \cite{Vallejos:2020} proposed an approach for assessing the agreement between two continuous responses when observations of both variables are georeferenced in space. For an increasing domain sampling scheme, they established the asymptotic normality of the sample spatial concordance coefficient for a bivariate Gaussian process with a Wendland covariance function. An image analysis example was used to illustrate both the strengths and limitations of the method. A generalization of the probability of agreement for geostatistical data was recently explored in \cite{Acosta:2024}. Additionally, concordance indices for measuring the correspondence between images have been applied in neuroimaging studies for comparing brain maps \citep{Alex:2018, Markello:2021}. The extension of agreement indices to lattice data, where observations are associated with areal units rather than point locations, remains an open problem.

This paper introduces a coefficient for measuring agreement between two lattice sequences observed in the same areal units. The motivation for this extension arises from the analysis of two different methodologies used to measure poverty rates in Chile. Our coefficient is a modified version of the multivariate coefficient studied by \cite{King:2007}, where the dependency of the multivariate lattice process is accounted for using a non-negative definite matrix of weights. To achieve this, we assume that the bivariate process follows a Multivariate Conditionally Autoregressive (GMCAR) process \citep{Jin:2005}. Due to the hierarchical structure, the inference is approached from a Bayesian perspective. Since the proposed agreement index is a function of the model parameters, inference is conducted using summaries from the Bayesian estimates. The methodology is illustrated through an analysis of poverty rates in Chile, with HPD intervals provided for the poverty rates in the Metropolitan and Valparaíso regions, the two largest regions in Chile.

We believe this work addresses a methodological gap in the understanding of agreement coefficients. The poverty rates enhance the applicability of the coefficient because numerous social variables are commonly measured in areal units, impacting the resources each areal unit (county) receives from the government.

The paper is structured as follows. In Section 2, we introduce the poverty dataset and the corresponding poverty rates. Section 3 presents the methodological aspects for constructing the concordance coefficient for lattice data. In Section 4, we briefly outline the Bayesian estimation method for the GMCAR process. Section 5 revisits the analysis of poverty rates and provides a statistical solution for the agreement problem between the Horvitz-Thompson (HT) and Small Area Estimation (SAE) coefficients, which have been calculated for every county in Chile. Finally, Section 6 outlines concluding comments, discussions, and future research directions.

\section{The poverty dataset}\label{sec:data}

\subsection{Poverty in Chile}
The eradication of poverty has been a central focus of public policies in Chile, guiding various efforts in this area. Estimates from nationwide surveys indicate, in the recent past \citep{Min:2012}, that the poverty rate has declined since the early 1990s, suggesting some progress towards this goal. However, erratic time series patterns have emerged in small counties, the smallest territorial units in Chile. For extremely small counties, poverty rate estimates are unavailable for certain periods due to the survey design, which traditionally prioritizes accurate estimates for larger regions. To effectively monitor trends, identify key factors, develop impactful public policies, and eliminate poverty at the county level, new methodologies have been introduced since 2011.

Chile’s primary source of poverty statistics is the National Socioeconomic Characterization Survey (CASEN), conducted by the Ministry of Social Development every 2 to 3 years since 1987. This survey collects samples from most counties. Socioeconomic data at the county level is crucial for formulating and evaluating public policies, particularly since municipalities serve as the first point of contact for Chileans with their local government.

Starting in 2007, regulations were introduced concerning the design of surveys and algorithms. In 2010, an expert committee was formed to provide recommendations for future innovations. They highlighted several concerns and suggestions for improving estimates, prompting the government to assemble a group of specialists tasked with developing a new methodology to enhance the precision of county-level estimates. This group recommended employing SAE methodology to assess poverty rates at the county level. The adoption of SAE represented a significant methodological advancement and has been employed in the CASEN survey ever since.

\subsection{Poverty rates and their estimates}
In Chile, poverty is assessed through the poverty rate, or Headcount Index, which indicates the share of households with incomes below a specified threshold \citep{Casas:2016}. Central to this measurement is the poverty line, which, in many Latin American countries, is based on the cost of a basic basket of essential food and non-food goods. This poverty line is expressed on a per capita basis\footnote{The methodology for determining Chile's poverty line was established by the Comisión Económica para América Latina y el Caribe (CEPAL) in 1990 and has remained largely unchanged since then
}.

In this paper, we analyzed the dataset from the CASEN survey conducted in 2011, which featured calculations for both the HT  and  SAE coefficients, along with various other social variables. The dataset includes poverty measures for 334 counties across Chile. Statistical analyses of the CASEN survey are documented in various publications. For more details, refer to \cite{Casas:2016}. Because we have specific interest in the spatial effects associated with the HT and SAE estimates, our analysis will be carried out mainly in the Matropolitan area (Santiago) where there are 52 counties,  and in the Valparaiso región which contains 34 counties.

Throughout this paper, we will denote the observed poverty rates from the Horvitz-Thompson and Small Area Estimation methods as HT and SAE, respectively, while also using the same notation to refer to the methods themselves. 

\begin{figure}
    \centering
    \includegraphics[width=1.0\linewidth]{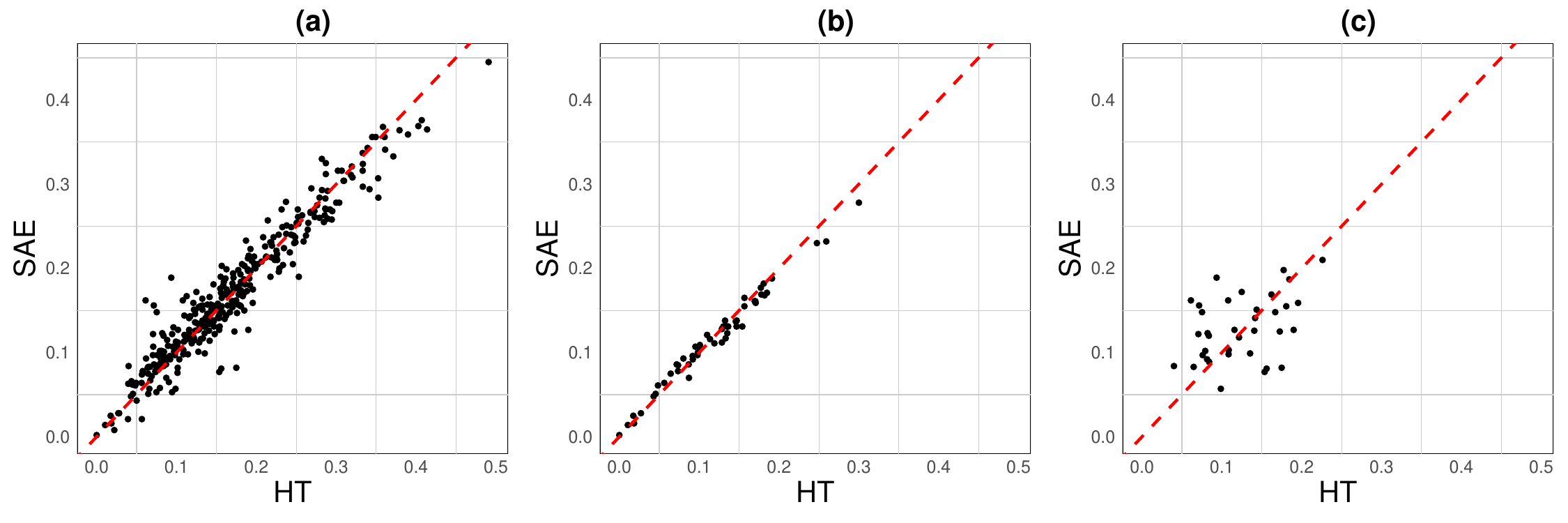}
    \caption{SAE versus HT. \textbf{(a)} Coefficients for the 334 counties in Chile; \textbf{(b)} Coefficients for the Metropolitan Region using 52 counties; \textbf{(c)} Coefficients for the Valparaiso Region using 38 counties.}
    \label{fig:Scatterplots}
\end{figure}

\begin{figure}
    \centering
    \includegraphics[width=1\linewidth]{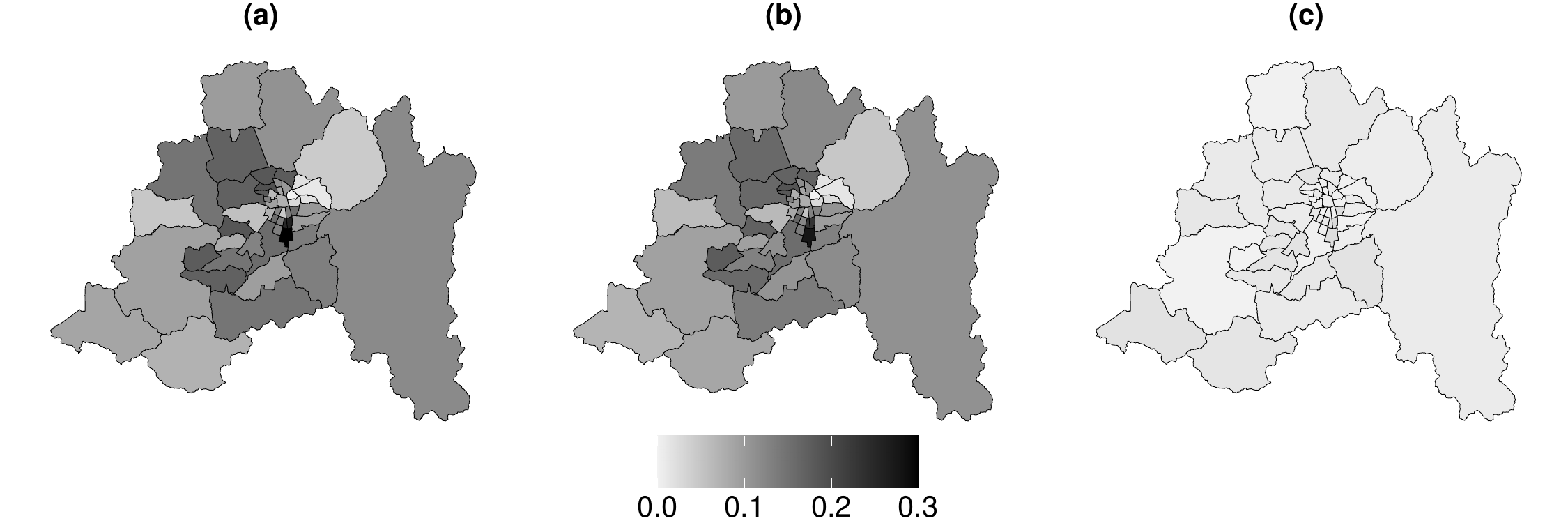}
    \caption{For the Metropolitan area: \textbf{ (a)} Poverty rates using the HT methodology; \textbf{(b)} Poverty rates the SAE methodology; \textbf{(c)} Absolute difference between the two rates.}
    \label{fig:RM_EDA}
\end{figure}

\begin{figure}
    \centering
    \includegraphics[width=1.0\linewidth]{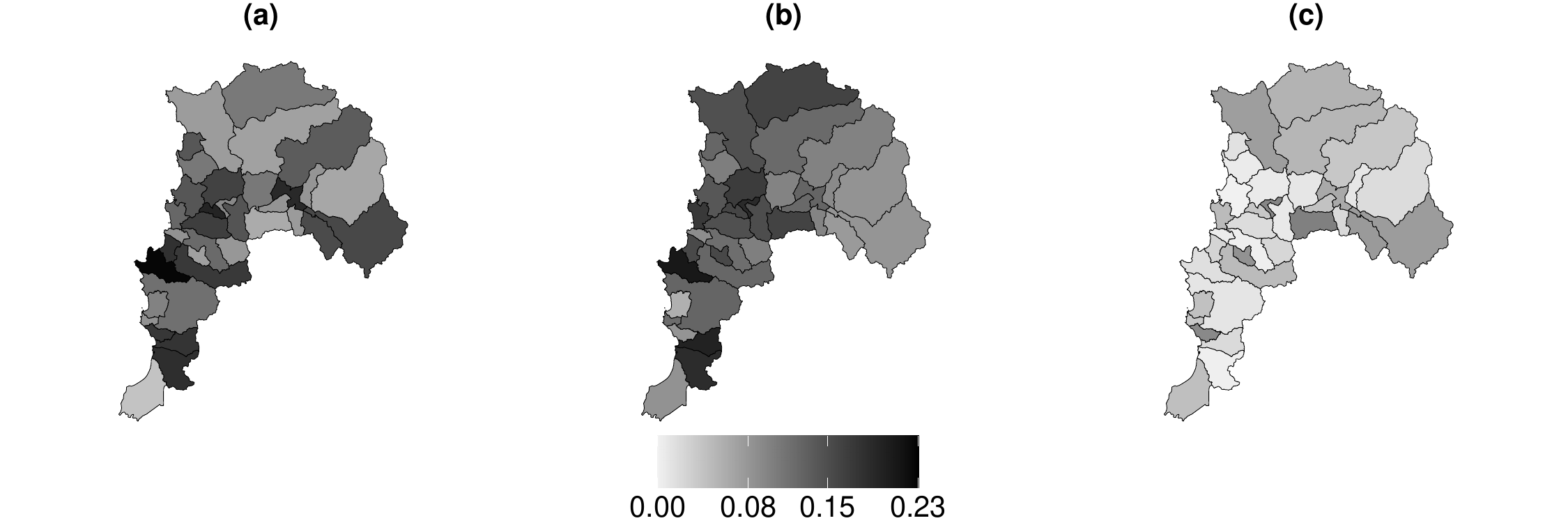}
    \caption{For the Valparaíso region: \textbf{(a)} Poverty rates using the HT methodology; \textbf{(b)} Poverty rates using the SAE methodology; \textbf{(c)} Absolute difference between the two rates.}
    \label{fig:V_EDA}
\end{figure}

Figure \ref{fig:Scatterplots} depicts scatterplots of the poverty rates obtained from the methods across Chile, the Metropolitan Area, and the Valparaíso region. Visually, there is a high level of agreement between the methods for both Chile as a whole and the Metropolitan Area. However, in the Valparaíso region, some discrepancies are observed, which merit closer attention. To visualize the spatial associations, Figures \ref{fig:RM_EDA} and \ref{fig:V_EDA} display the poverty rates for the Metropolitan area and Valparaíso region, respectively. In each case, the maps shown in (c) represent the differences between the maps in (a) and (b). For the Metropolitan area, this difference map exhibits less structure compared to the Valparaíso region, highlighting a more pronounced discrepancy between the rate estimates for Valparaíso. This is an indication that the spatial structure plays a fundamental role and has to be taken into account in the developed concordance indices.

The objective of this paper is to offer a statistical solution for quantifying the agreement between two lattice sequences. The solution presented in the following sections is model-based, rooted in the framework of bivariate spatial processes

\section{Concordance for Lattice Data}\label{sec:lattice}

In this section, we briefly describe agreement coefficients for continuous variables and their generalization to the multivariate case. We also review typical multivariate autoregressive processes defined on lattices. Finally, we introduce a new agreement coefficient for areal data.

\subsection{Agreement coefficients}

The concordance correlation coefficient, first introduced by \cite{Lin:1989} , evaluates the degree of agreement between two continuous measurements. Let $(X_{11},X_{21})^{\top}$, \ldots, $(X_{1n},X_{2n})^{\top}$ be pairs of samples independently selected 
from a bivariate normal population with
\begin{equation*}
    \boldsymbol{\mu}=\begin{pmatrix}
    \mu_1 \\
    \mu_1 
    \end{pmatrix}\in\mathbb{R}^2\quad\mbox{and}\quad \boldsymbol{\Sigma}=\begin{pmatrix}
    \sigma_{1}^2 & \sigma_{12}\\
    \sigma_{21} & \sigma_{2}^2
    \end{pmatrix}\in\mathbb{R}^{2\times 2}.
\end{equation*}
The degree of concordance between $X_1$ and $X_2$ can be characterized by the expected value 
of the squared difference, i.e.,
\begin{equation*}
\mathbb{E}[\left(X_1-X_2\right)^2] = \left(\mu_1-\mu_2\right)^2+\left(\sigma_1-\sigma_2\right)^2+2(1-\rho) \sigma_1 \sigma_2,
\end{equation*}
where $\rho=\sigma_{12}/(\sigma_1\sigma_2)$ is the Pearson's correlation coefficient.
If each pair, $(X_1,X_2)$, in the population are in perfect agreement, $\mathbb{E}[\left(X_1-X_2\right)^2]$ would 
be $0$. The following transformation is proposed to scale the index value between $-1$ and $1$:
\begin{equation*}
    \rho_c=1-\frac{\mathbb{E}[(X_1-X_2)^2]}{\mathbb{E}[(X_1-X_2)^2\mid\sigma_{12}=0]}=\frac{2 \sigma_{12}}{\sigma_1^2+\sigma_2^2+\left(\mu_1-\mu_2\right)^2}=\rho C,
\end{equation*}
where $C=\{(v^2+1/v+u^2)/2\}^{-1}$, $v=\sigma_1/\sigma_2$ y $u=(\mu_1-\mu_2)/\sqrt{\sigma_1\sigma_2}$. Here, $0 < C\leq 1$ is a bias correction factor. 

From the definition it is clear that $\rho_c$ posses the following properties 

\begin{enumerate}
    \item[i)] $-1\leq \rho_c\leq 1$.
    \item[ii)] $\rho_c=0$ if and only if $\rho=0$.
    \item[iii)] $\rho_c=\rho$ if and only if $\sigma_1=\sigma_2$ and $\mu_1=\mu_2$
    \item[iv)] $\rho_c=\pm 1$ if and only if $\rho=\pm 1$, $\mu_1=\mu_2$ y $\sigma_1=\sigma_2$.
\end{enumerate}

Inference on $\rho_c$ was developed using the Fisher Z-transformation and the Delta method, establishing the asymptotic normality of the estimated coefficient.

To generalize Lin's coefficient for lattice data to analyze poverty in Chile, we first introduce a multivariate extension that mirrors Lin's approach for random vectors.

Let $(\boldsymbol{X}_1,\boldsymbol{X}_2)^{\top}$ be a $2n \times 1$ Gaussian random vector from a $2n$-variate distribution with 
\begin{equation*}
    \boldsymbol{\mu}=\begin{pmatrix}
    \boldsymbol{\mu}_1 \\
    \boldsymbol{\mu}_2 
    \end{pmatrix}\in\mathbb{R}^{2n\times 1},\quad\quad \boldsymbol{\Sigma}=\begin{pmatrix}
    \boldsymbol{\Sigma}_{11} & \boldsymbol{\Sigma}_{12}\\
    \boldsymbol{\Sigma}_{12}^{\top} & \boldsymbol{\Sigma}_{22}
    \end{pmatrix}\in\mathbb{R}^{2n\times2n},
\end{equation*}
where $\boldsymbol{X}_1$ and $\boldsymbol{X}_2$ are the measurements of each method.

To characterize the level of agreement between $\boldsymbol{X}_1$ and $\boldsymbol{X}_2$, \cite{King:2007} consider the following expression:
\begin{equation*}
    \mathbb{E}[(\boldsymbol{X}_1-\boldsymbol{X}_2)^{\top}\boldsymbol{D}(\boldsymbol{X}_1-\boldsymbol{X}_2)]=\operatorname{Tr}\{\boldsymbol{D}(\boldsymbol{\Sigma}_{11}+\boldsymbol{\Sigma}_{22}-\boldsymbol{\Sigma}_{12}-\boldsymbol{\Sigma}_{12}^{\top})\}+\left(\boldsymbol{\mu}_1-\boldsymbol{\mu}_2\right)^{\top} \boldsymbol{D}\left(\boldsymbol{\mu}_1-\boldsymbol{\mu}_2\right),
\end{equation*}
where $\boldsymbol{D}\in\mathbb{R}^{n\times n}$ is a non-negative definite matrix matrix of weights between the different repeated measurements. Extending Lin's coefficient, we develop a repeated measures concordance coefficient defined through
\begin{align*}
    \varrho_c  &= 1 - \frac{\mathbb{E}[(\boldsymbol{X}_1 - \boldsymbol{X}_2)^{\top} \boldsymbol{D}(\boldsymbol{X}_1 - \boldsymbol{X}_2)]}{\mathbb{E}[(\boldsymbol{X}_1 - \boldsymbol{X}_2)^{\top} \boldsymbol{D}(\boldsymbol{X}_1 - \boldsymbol{X}_2)\mid\boldsymbol{\Sigma}_{12}=\boldsymbol{0}]}\\
    &= \frac{\operatorname{Tr}[\boldsymbol{D} \boldsymbol{\Sigma}_{12} + \boldsymbol{D} \boldsymbol{\Sigma}_{12}^{\top}]}{\operatorname{Tr}[\boldsymbol{D} \boldsymbol{\Sigma}_{11} + \boldsymbol{D} \boldsymbol{\Sigma}_{22}] + \left(\boldsymbol{\mu}_1 - \boldsymbol{\mu}_2\right)^{\top} \boldsymbol{D}\left(\boldsymbol{\mu}_1 - \boldsymbol{\mu}_2\right)}.
\end{align*}
It is noteworthy that $\varrho_c=\rho_c$ when $n=1$, and it possesses the following properties:
\begin{enumerate}
    \item[i)] $-1\leq\varrho_c\leq 1$.
    \item[ii)] $\varrho_c = 1$ if $\boldsymbol{X}_1=\boldsymbol{X}_2$ with probability $1$.
    \item[iii)] $\varrho_c = -1$ if $\boldsymbol{X}_1=-\boldsymbol{X}_2$ with probability $1$.
    \item[iv)] $\varrho_c = 0$ if there is no agreement between $\boldsymbol{X}_1$ and $\boldsymbol{X}_2$.
\end{enumerate}
Furthermore, inference on $\varrho_c$ is developed within the framework of $U$-statistics.

\subsection{Autoregressive models defined on lattices}

Let $S\subset \mathbb{R}^2$, and consider a finite partition of $S$ represented by the set $\{C_1,\ldots, C_n\}$, i.e., $S=\bigcup_{i=1}^{n} C_i$, such that $C_i\cap C_j=\emptyset$ for all $i\neq j$. Lattice data represent observations associated with each set in the partition. 

An established approach for modeling this type of data involves treating the value associated with lattice unit $C_i$ as conditional on a linear combination of values from neighboring units. The strength of this relationship is expected to increase with spatial proximity, forming the foundation of the Conditional Autoregressive (CAR) process \citep{Besag:1974, Cressie:1993}. Specifically, let $X_i$ denote the measurement for unit $C_i$, $i=1,\ldots, n$. A CAR process for these observations is defined by the full conditional distribution
\begin{equation}
\label{eqn:modelCAR}
X_i \mid \boldsymbol{X}_{-(i)}\sim\operatorname{N}\hspace{-2pt}\left(\rho\sum_{j\in \partial_i} b_{ij}x_j,\tau_i^{-1}\right),\quad i=1,\ldots,n,
\end{equation}
where $j\in \partial_i$ denotes those indices whose lattice unit $C_j$ shares boundary with lattice unit $C_i$, $b_{ij}$ are spatial weights, $\rho$ is a spatial autocorrelation parameter and $\tau_i$ is the conditional variance.

From the Hammersley-Clifford Theorem and Brook's Lemma, the full conditional distributions in (\ref{eqn:modelCAR}) uniquely determine the joint distribution,
\begin{equation*}
    \boldsymbol{X} \sim \operatorname{N}\hspace{-2pt}\left(\boldsymbol{0}_n,\left[\boldsymbol{D}_\tau(\boldsymbol{I}_n-\rho \boldsymbol{B})\right]^{-1}\right),
\end{equation*}
where $\boldsymbol{B}=\{b_{ij}\}_{i,j=1}^n\in\mathbb{R}^{n\times n}$ satisfying $b_{i i}=0$, and $\boldsymbol{D}_\tau=\operatorname{diag}\{\tau_1,\ldots,\tau_n\}$. 

Usually, a contiguity matrix $\boldsymbol{W}_1=\{w_{ij}\}_{i,j=1}^n\in\mathbb{R}^{n\times n}$ is defined such that $w_{ij}=1$ if $C_i$ and $C_j$ share some common boundary, otherwise $w_{ij}=0$.  Additionally, it is standard to set $w_{ii}=0$. Next, we define the matrix of total neighbords for each lattice unit, as $\boldsymbol{D}_{w}=\operatorname{diag}\{w_{1+},\ldots,w_{n+}\}\in\mathbb{R}^{n\times n}$ where
$w_{i+}=\sum_{j\neq i}w_{ij}$ denotes the number of lattice units sharing a boundary with $C_i$. These definitions are commonly used to determine the typical forms of the matrices $\boldsymbol{B}$ and $\boldsymbol{D}_{\tau}$, where we set $\boldsymbol{D}_{\tau} = \tau \boldsymbol{D}_{w}$ and $\boldsymbol{B}=\boldsymbol{D}_{w}^{-1}\boldsymbol{W}_1$, thereby providing the full specification of the CAR process.

Several generalizations of multivariate CAR processes have been introduced in the literature \citep{Mardia:1988, Kim:2001, Gelfand:2003, Jin:2005, Jin:2007, Greco:2009, MartinezBeneito:2013, Cressie:2016}. In this study, we adopt the Generalized Multivariate CAR model proposed by \cite{Jin:2005} for its computational efficiency and its seamless extension from univariate CAR processes.

Consider the zero mean joint distribution
of $\boldsymbol{X}_1$ and $\boldsymbol{X}_2$ as
\begin{equation}\label{bivariate}
    \begin{pmatrix}
    \boldsymbol{X}_1 \\
    \boldsymbol{X}_2 
\end{pmatrix}\sim \mathrm{N}\left(\begin{pmatrix}
    \boldsymbol{0}_n \\
    \boldsymbol{0}_n 
    \end{pmatrix},\begin{pmatrix}
\boldsymbol{\Sigma}_{11} & \boldsymbol{\Sigma}_{12} \\
\boldsymbol{\Sigma}_{12}^{\top} & \boldsymbol{\Sigma}_{22}
\end{pmatrix}\right).
\end{equation}
From standard multivariate normal theory, we have
\begin{equation*}
    \mathbb{E}[\boldsymbol{X}_1|\boldsymbol{X}_2]=\boldsymbol{\Sigma}_{11}\boldsymbol{\Sigma}_{22}^{-1}\boldsymbol{X}_2\quad\mbox{and}\quad\mathbb{V}\mathrm{ar}[\boldsymbol{X}_1|\boldsymbol{X}_2]=\boldsymbol{\Sigma}_{11\cdot 2}=\boldsymbol{\Sigma}_{11}-\boldsymbol{\Sigma}_{12}\boldsymbol{\Sigma}_{22}^{-1}\boldsymbol{\Sigma}_{12}^{\top}.    
\end{equation*}
Defining $\boldsymbol{A}=\boldsymbol{\Sigma}_{12}\boldsymbol{\Sigma}_{22}^{-1}$, the covariance matrices of $\boldsymbol X_1$ and $\boldsymbol X_2$ can be expressed respectively as $\boldsymbol{\Sigma}_{11}=\boldsymbol{\Sigma}_{11\cdot 2}+\boldsymbol{A}\boldsymbol{\Sigma}_{22}\boldsymbol{A}^{\top}$ and $\boldsymbol{\Sigma}_{12}=\boldsymbol{A}\boldsymbol{\Sigma}_{22}$. Since $\boldsymbol{X}_1|\boldsymbol{X}_2\sim \mathrm{N}(\boldsymbol{A}\boldsymbol{X}_2,\boldsymbol{\Sigma}_{11\cdot2})$ and $\boldsymbol{X}_2\sim\mathrm{N}(\boldsymbol{0}_n,\boldsymbol{\Sigma}_{22})$, joint distribution of $\boldsymbol X=(\boldsymbol X_1 ^{\top} \boldsymbol X_2 ^{\top})^{\top}$
is $f(\boldsymbol{X})=f(\boldsymbol{X}_1|\boldsymbol{X}_2)f(\boldsymbol{X}_2)$.

Following univariate CAR theory, \cite{Jin:2005} assume that $\boldsymbol{\Sigma}_{11\cdot 2}=\left[\left(\boldsymbol{D}_{w}-\rho_1 \boldsymbol{W}_1\right) \tau_1\right]^{-1}$ and $\boldsymbol{\Sigma}_{22}=\left[\left(\boldsymbol{D}_{w}-\rho_2 \boldsymbol{W}_1\right) \tau_2\right]^{-1}$, where $\rho_1$ and $\rho_2$ are the smoothing parameters associated with the conditional distribution $\boldsymbol{X}_1|\boldsymbol{X}_2$ and the marginal distribution $\boldsymbol{X}_2$ respectively, and $\tau_1$ and $\tau_2$ scale the precision of $\boldsymbol{X}_1|\boldsymbol{X}_2$ and $\boldsymbol{X}_2$, respectively. The resulting joint distribution remains proper as long as these two CAR distributions are valid, requiring only that $|\rho_1|<1$ and $|\rho_2|<1$ to ensure the positive definiteness of the covariance matrix in (\ref{bivariate}).

Regarding the $\boldsymbol{A}$ matrix, we propose a linking form given by $\boldsymbol{A}=\eta_0\boldsymbol{I}_{n}+\eta_1\boldsymbol{W}_1$, where $\eta_0$ and $\eta_1$ are bridging parameters that associate $X_{1i}$ with $X_{2i}$ and $X_{2j}$ for $j \neq i$.

\begin{figure}
    \centering
    \includegraphics[width=0.9\linewidth]{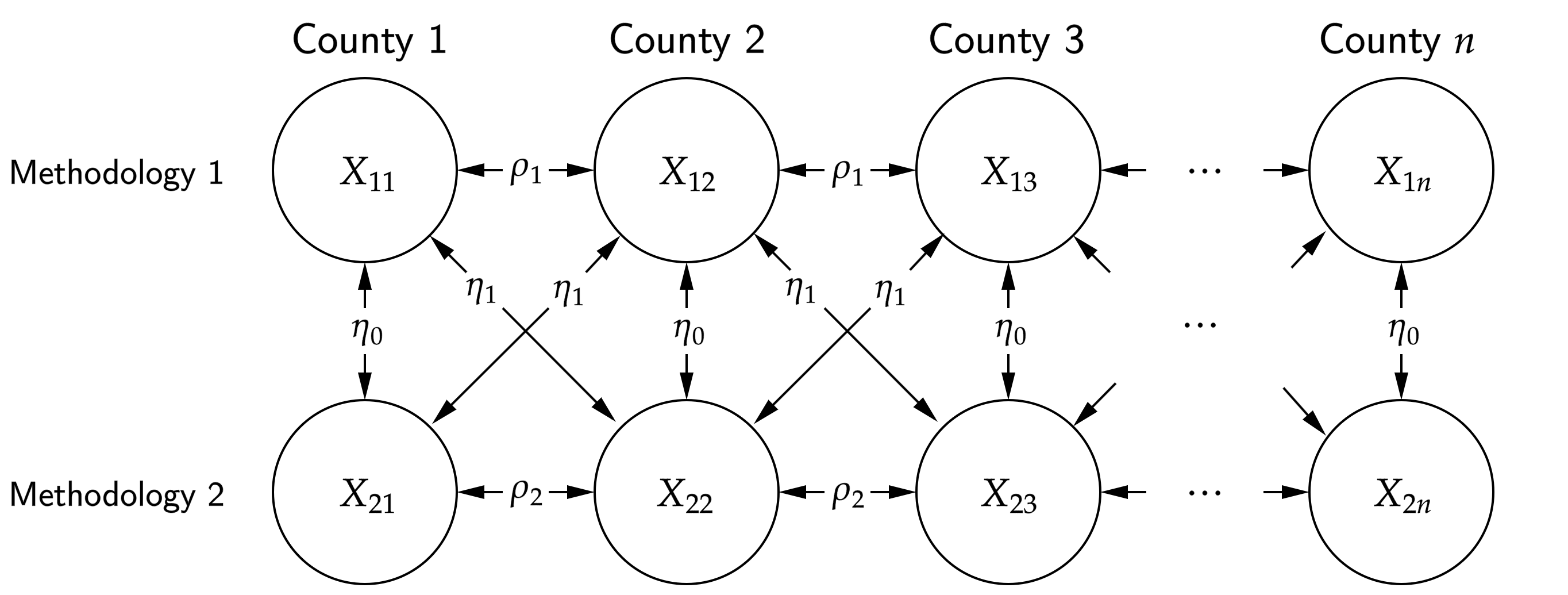}
    \caption{Dependence representation for a bivariate GMCAR process.}
    \label{fig:GMCAR}
\end{figure}

As a consequence, the blocks of the covariance matrix in the joint distribution of $\boldsymbol X$ can be written as
\begin{align*}
    &\boldsymbol{\Sigma}_{11}=\left[\tau_1\left(\boldsymbol{D}_{w}-\rho_1 \boldsymbol{W}_1\right)\right]^{-1}+\left(\eta_0 \boldsymbol{I}_{n}+\eta_1 \boldsymbol{W}_1\right)\left[\tau_2\left(\boldsymbol{D}_w-\rho_2 \boldsymbol{W}_1\right)\right]^{-1}\left(\eta_0 \boldsymbol{I}_{n}+\eta_1 \boldsymbol{W}_1\right),\\
    &\boldsymbol{\Sigma}_{12}=\left(\eta_0 \boldsymbol{I}_{n}+\eta_1 \boldsymbol{W}_1\right)\left[\tau_2\left(\boldsymbol{D}_w-\rho_2 \boldsymbol{W}_1\right)\right]^{-1},\\
    &\boldsymbol{\Sigma}_{22}=\left[\tau_2\left(\boldsymbol{D}_w-\rho_2 \boldsymbol{W}_1\right)\right]^{-1}.
\end{align*}
This model is denoted as $\mathrm{GMCAR}(\rho_1,\rho_2,\eta_0,\eta_1,\tau_1,\tau_2)$. Figure \ref{fig:GMCAR}  presents a visual representation of the spatial autocorrelation parameters and the bridging parameters, illustrating their roles in the relationship between spatial units and how they affect the dependence structure across different regions.

\subsection{A new coefficient}

To define the spatial concordance coefficient for lattice data, we apply the generalized concordance coefficient developed by \cite{King:2007}, assuming that the lattice data can be effectively represented by a $\mathrm{GMCAR}$ 
 model.

\begin{definition}
    Let $\boldsymbol{X}=(\boldsymbol{X}_1^{\top},\boldsymbol{X}_2^{\top})^{\top}$ be a $2n\times 1$ random vector from a $2n-$variate Gaussian distribution. Assume that $\boldsymbol X$  follows a $\mathrm{GMCAR}(\rho_1,\rho_2,\eta_0,\eta_1,\tau_1,\tau_2)$ model, where $\boldsymbol{X}_i=(X_{i1},\ldots, X_{in})^{\top}$ is a $n\times 1$ random vector. That is,
    $$
    \begin{pmatrix}
        \boldsymbol{X}_1 \\
        \boldsymbol{X}_2 
    \end{pmatrix} \sim \mathrm{N}\hspace{-2pt}\left(\begin{pmatrix}
        \boldsymbol{\mu}_{1} \\
        \boldsymbol{\mu}_{2}
        \end{pmatrix},\begin{pmatrix}
    \boldsymbol{\Sigma}_{11} & \boldsymbol{\Sigma}_{12} \\
    \boldsymbol{\Sigma}_{12}^{\top} & \boldsymbol{\Sigma}_{22}
    \end{pmatrix}\right),
    $$
    where $\mathbb{E}[\boldsymbol X_i]=\boldsymbol{\mu}_{i}$,  $i=1,2$, and
    \begin{align*}
    &\boldsymbol{\Sigma}_{11}=\left[\tau_1\left(\boldsymbol{D}_w-\rho_1 \boldsymbol{W}_1\right)\right]^{-1}+\left(\eta_0 \boldsymbol{I}_{n}+\eta_1 \boldsymbol{W}_1\right)\left[\tau_2\left(\boldsymbol{D}_w-\rho_2 \boldsymbol{W}_1\right)\right]^{-1}\left(\eta_0 \boldsymbol{I}_{n}+\eta_1 \boldsymbol{W}_1\right),\\
    &\boldsymbol{\Sigma}_{12}=\left(\eta_0 \boldsymbol{I}_{n}+\eta_1 \boldsymbol{W}_1\right)\left[\tau_2\left(\boldsymbol{D}_w-\rho_2 \boldsymbol{W}_1\right)\right]^{-1},\\
    &\boldsymbol{\Sigma}_{22}=\left[\tau_2\left(\boldsymbol{D}_w-\rho_2 \boldsymbol{W}_1\right)\right]^{-1}.
    \end{align*}
    The spatial concordance coefficient, $\varrho_{s,c}$, is defined as
    \begin{equation}\label{lattice_coef}
    \varrho_{s,c}=\frac{\operatorname{Tr}[\boldsymbol{J}_n \boldsymbol{\Sigma}_{12}+\boldsymbol{J}_n \boldsymbol{\Sigma}_{12}^{\top}]}{\operatorname{Tr}[\boldsymbol{J}_n \boldsymbol{\Sigma}_{11}+\boldsymbol{J}_n \boldsymbol{\Sigma}_{22}]+\left(\boldsymbol{\mu}_1-\boldsymbol{\mu}_2\right)^{\top} \boldsymbol{J}_n\left(\boldsymbol{\mu}_1-\boldsymbol{\mu}_2\right)},
    \end{equation}
    where $\boldsymbol{J}_n=\boldsymbol{1}_n\boldsymbol{1}_n^{\top}$.
\end{definition}

In \eqref{lattice_coef}, we set $\boldsymbol{D}=\boldsymbol{J}_n$ because the estimated covariance matrix already incorporates the inter- and intra-observation weights through spatial autocorrelation and linking parameters. Additionally, the spatial concordance coefficient naturally extends Lin's multivariate coefficient by incorporating lattice dependence into the covariance structure. Furthermore, as noted by \cite{Jin:2005}, the coefficient can be generalized to include dependencies among higher-order neighbors via the matrix  $\boldsymbol{A}$, thereby increasing the number of linking parameters as follows:
\begin{equation*}
    \boldsymbol{A}=\eta_0 \boldsymbol{I}_n+\sum_{j=1}^{n-1} \eta_j \boldsymbol{W}_j,
\end{equation*}
where $\boldsymbol{W}_j$ denotes de contiguity matrix for $j$-order neighbors. This offers practitioners a flexible approach for measuring concordance.

\section{Bayesian inference}\label{sec:Bayesian}

Statistical inference for the sample coefficient associated with (\ref{lattice_coef}) was approached from a Bayesian perspective. While the asymptotic distribution of similar coefficients has been studied using a classical approach with an increasing domain sampling scheme \citep{Vallejos:2020}, these assumptions cannot be easily extended to areal processes. On the other hand, most areal models based on $\mathrm{CAR}$ processes are typically fitted within a hierarchical framework.

Assume that $\boldsymbol X$ follows a  $\mathrm{GMCAR}(\rho_1,\rho_2,\eta_0,\eta_1,\tau_1,\tau_2)$ process. Then,
\begin{equation*}
    \boldsymbol{X}_1| \boldsymbol{X}_2 \sim  \mathrm{N}\left(\boldsymbol{\mu}_1+\left(\eta_0 \boldsymbol{I}_{n}+\eta_1 \boldsymbol{W}_1\right)\left(\boldsymbol{X}_2- \boldsymbol{\mu}_2\right),\left[\tau_1\left(\boldsymbol{D}_w-\rho_1 \boldsymbol{W}_1\right)\right]^{-1}\right),
\end{equation*}
and $\boldsymbol{X}_2 \sim \mathrm{N}\left(\boldsymbol{\mu}_2,\left[\tau_2(\boldsymbol{D}_w-\rho_2 \boldsymbol{W}_1\right)\right]^{-1})$,
therefore,  the joint distribution of $\boldsymbol{X}$ is given by
\begin{align*}
 f(\boldsymbol{X}| \boldsymbol{\mu}, \boldsymbol{\tau}, \boldsymbol{\rho}, \boldsymbol{\eta}) &\propto \tau_1^{n / 2}\operatorname{det}(\boldsymbol{D}_w-\rho_1 \boldsymbol{W}_1)^{1 / 2} \\
& \times \exp \left\{-\frac{\tau_1}{2}\left[\boldsymbol{X}_1-\boldsymbol{\mu}_1-\left(\eta_0 \boldsymbol{I}_{n}+\eta_1 \boldsymbol{W}_1\right)\left(\boldsymbol{X}_2-\boldsymbol{\mu}_2\right)\right]^{\top}\right. \\
& \times\left(\boldsymbol{D}_w-\rho_1 \boldsymbol{W}_1\right)\left[\boldsymbol{X}_1-\boldsymbol{\mu}_1-\left(\eta_0 \boldsymbol{I}_{n}+\eta_1 \boldsymbol{W}_1\right)\left(\boldsymbol{X}_2- \boldsymbol{\mu}_2\right)\right]\Big\} \\
& \times \tau_2^{n / 2}\operatorname{det}(\boldsymbol{D}_w-\rho_2 \boldsymbol{W}_1)^{1 / 2} \\
& \times \exp \left\{-\frac{\tau_2}{2}\left(\boldsymbol{X}_2-\boldsymbol{\mu}_2 \right)^{\top}\left(\boldsymbol{D}_w-\rho_2 \boldsymbol{W}_1\right)\left(\boldsymbol{X}_2-\boldsymbol{\mu}_2 \right)\right\},
\end{align*}
where $\boldsymbol{\mu}=(\boldsymbol{\mu}_1^{\top},\boldsymbol{\mu}_2^{\top})^{\top}$, $\boldsymbol{\tau}=(\tau_1,\tau_2)^{\top}$, $\boldsymbol{\eta}=(\eta_0,\eta_1)^{\top}$ and $\boldsymbol{\rho}=(\rho_1,\rho_2)^{\top}$. Thus, the posterior distribution is 
\begin{equation*}
    f(\boldsymbol{\mu}, \boldsymbol{\tau}, \boldsymbol{\rho}, \boldsymbol{\eta}| \boldsymbol{X})\propto f(\boldsymbol{X}| \boldsymbol{\mu}, \boldsymbol{\tau}, \boldsymbol{\rho}, \boldsymbol{\eta}) f(\boldsymbol{\mu}) f(\boldsymbol{\tau}) f(\boldsymbol{\rho}) f(\boldsymbol{\eta}),
\end{equation*}
assuming independent priors. The parameters are sampled from the posterior distribution using MCMC, including the Metropolis random walk method, as implemented in the \texttt{rjags} package within the $\mathrm{R}$ programming language \citep{RCoreTeam:2024, Plummer:2024}. Then, the distribution of the spatial concordance coefficient \eqref{lattice_coef} is constructed as a plug-in type estimator, 
\begin{equation}\label{bay_est}
\widehat{\varrho}_{s,c}=\varrho_{s,c}(\widehat{\boldsymbol{\theta}}),
\end{equation}
where $\boldsymbol{\theta}=(\boldsymbol{\mu},\boldsymbol{\tau},\boldsymbol{\eta},\boldsymbol{\rho})^{\top}$.

\section{An application}\label{sec:application}

For the poverty dataset, we assume that $\boldsymbol{\mu}_i = \mu_i \boldsymbol{1}_n, i=1,2$. This assumption implies a constant mean for each methodology. In this context, the non-spatial component and the inherent structure of the data are incorporated into the mean, while the spatial characteristics of the regions associated with each observation are captured through the covariance matrix.

We define $X_{ij} = \mu_i + \varepsilon_{ij}+\phi_{ij}$, where $X_{ij}$ represents the poverty rate for the $i$-th methodology in the $j$-th lattice unit (e.g., counties in the Valparaiso or Metropolitan regions). Here, $\mu_i$ denotes the mean of the $i$-th methodology, $\varepsilon_{ij}\sim\mathrm{N}(0,\sigma_{i})$ is noise term for each methodology, and $\phi_{ij}$ is a spatial disturbance term, modeled as $\mathrm{GMCAR}(\rho_1, \rho_2, \eta_0, \eta_1, \tau_1, \tau_2)$, which captures the spatial dependence across the lattice units. In this case, we denote $\boldsymbol{X}_1$ as SAE, such that $\boldsymbol{X}_1 = \text{SAE} \mid \boldsymbol{X}_2 = \text{HT}$ makes sense within the established hierarchical structure of the model. This is because SAE represents the new methodology, while HT represents the existing one (please refer to \ref{sec: Appendix A} for additional information ).

In both regions, we adjust the GMCAR process using three configurations based on first, second, and third-order neighbors:
\begin{enumerate}
    \item[(a)] $\boldsymbol{A} = \eta_0 \boldsymbol{I}_n + \eta_1 \boldsymbol{W}_1$.
    \item[(b)] $\boldsymbol{A} = \eta_0 \boldsymbol{I}_n + \eta_1 \boldsymbol{W}_1+\eta_2 \boldsymbol{W}_2$.
    \item[(c)] $\boldsymbol{A} = \eta_0 \boldsymbol{I}_n + \eta_1 \boldsymbol{W}_1+\eta_2 \boldsymbol{W}_2+\eta_3\boldsymbol{W}_3$.
\end{enumerate}
We then select the model with the lowest Deviance Information Criterion (DIC) value and compute  $\widehat{\rho}_{s,c}$ for the selected model. Furthermore, the priors in these three cases were specified as follows: $\rho_i \sim \mathrm{Unif}(0, 1)$, $\tau_i \sim \mathrm{Gamma}(0.1, 0.1)$, $\sigma_i \sim \mathrm{Gamma}(0.1, 0.1)$, $\eta_0 \sim \mathrm{N}(0, 100)$, $\eta_1 \sim \mathrm{N}(0, 100)$. For the Santiago Metropolitan region, the prior distribution for the mean was
$(\boldsymbol{\mu}_1^{\top}, \boldsymbol{\mu}_2^{\top})^{\top} \sim \mathrm{N}(0.118 \boldsymbol{1}_{2n}, 10 \boldsymbol{I}_{2n}),$
whereas for the Valparaiso region, 
$
(\boldsymbol{\mu}_1^{\top}, \boldsymbol{\mu}_2^{\top})^{\top} \sim \mathrm{N}(0.126 \boldsymbol{1}_{2n}, 10 \boldsymbol{I}_{2n}).
$
The MCMC parameters included a single chain with a total of 30,000 iterations, of which the first 15,000 were discarded as burn-in to ensure convergence. The samples were retained without thinning, resulting in 15,000 posterior samples for inference.

\begin{figure}
    \centering
    \begin{minipage}{0.5\textwidth}
        \centering
        \includegraphics[width=\textwidth]{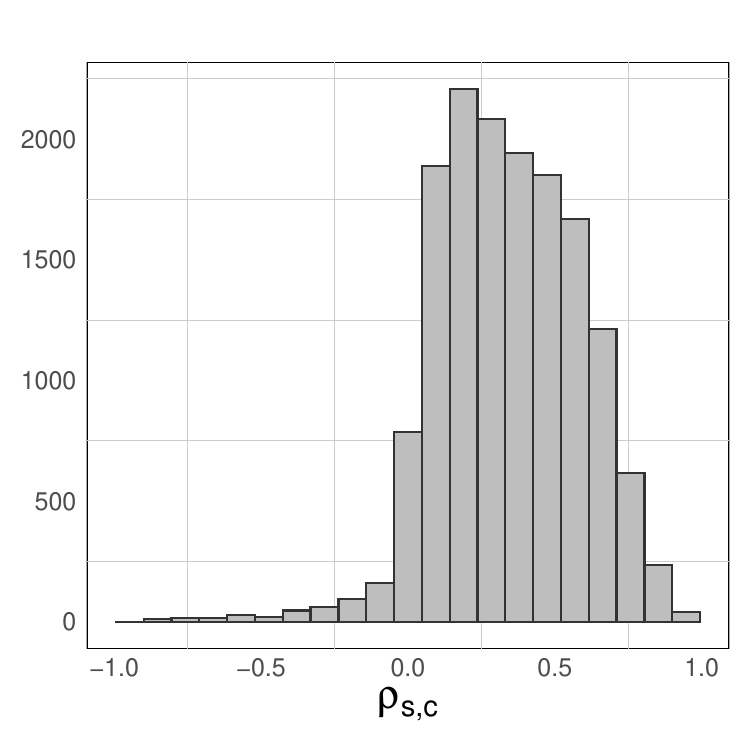}
    \end{minipage}\hfill
    \begin{minipage}{0.5\textwidth}
        \centering
        \includegraphics[width=\textwidth]{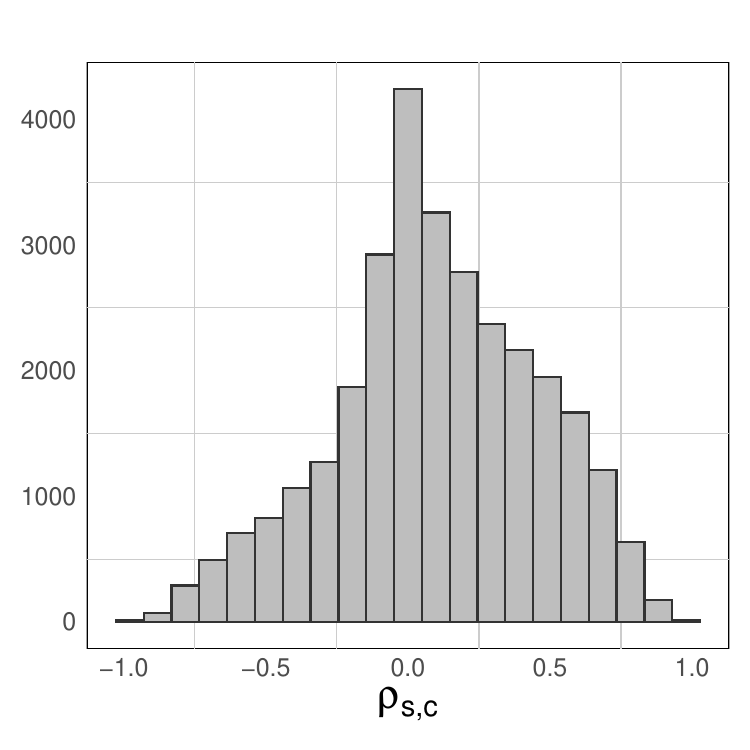}
    \end{minipage}
    \caption{Posterior distribution of the spatial concordance coefficient $\rho_{s,c}$ in the application for the Santiago Metropolitan Region (\textbf{Left}) and Valparaiso Region (\textbf{Right}).}
    \label{fig:rho_sc_plots}
\end{figure}

In both regions, model (a) provided the best fit based on the DIC values shown in Table \ref{tab:DIC_values}. The parameters related to the spatial and non-spatial components are detailed in \ref{sec: Appendix B}.  The distributions of $\widehat{\rho}_{s,c}$ based on samples from the posteriors are displayed in Figure \ref{fig:rho_sc_plots}, while the posterior means and $95\%$ HPD intervals are presented in Table \ref{tab:rho_sc_summary}.

\begin{table}[H]
    \centering
    \caption{DIC for three models for each region.}
    \begin{tabular}{cccc}
    \toprule 
     & (a) & (b) & (c) \\ \midrule
    Santiago & $-128.9$ & $-128.1$ & $-127.9$ \\
    & & & \\ 
    Valparaíso & $-77.0$ & $-66.2$ & $-61.4$ \\
    \bottomrule
    \end{tabular}
    \label{tab:DIC_values}
\end{table}

\begin{table}[H]
    \centering
    \caption{Posterior mean of the spatial concordance coefficient with $95\%$ HPD intervals.}
    \begin{tabular}{cc}
    \toprule 
     & $\widehat{\rho}_{s,c}$  \\ 
     \midrule
     Santiago & $0.343$ \\ \vspace{0.5cm}
     & $(-0.047, 0.799)$ \\
     Valparaíso & $0.059$ \\
     & $(-0.629, 0.749)$  \\
    \bottomrule
    \end{tabular}
    \label{tab:rho_sc_summary}
\end{table}

On one hand, these results provide a statistical solution to the problem of quantifying the agreement between two lattice sequences on the plane. On the other hand, they indicate that both methodologies show a certain level of concordance, consistent with the intuitive visual appeal of Figure \ref{fig:Scatterplots}. However, for the Valparaíso region, a lack of agreement is evident from the results shown in Table \ref{tab:rho_sc_summary}. Since both methodologies were applied solely to the 2011 survey, further analysis of the factors underlying these discrepancies is challenging.

The codes developed in \texttt{R} for the application presented in this section are available in the repository \url{https://github.com/clementeferrer/Code-for-Lattice-Concordance}.

\section{Discussion and final comments}
This paper introduced a new coefficient for evaluating the agreement between two lattice sequences measured at the same sites. The proposed generalization was motivated by poverty studies conducted at the county level in Chile, where a well-designed internal survey measuring various social aspects was administered every two years. We view this effort as the starting point for several future studies that could be conducted within a similar framework. The present research offers valuable insights from both modeling and practical perspectives, which can be applied in future explorations.

To capture both autocorrelation and cross-correlation between spatial components, the GMCAR model offers a flexible and computationally efficient framework for analyzing spatially structured data. It extends the univariate CAR approach to a multivariate setting, enabling the modeling of correlated spatial variables across a lattice while effectively capturing local spatial dependencies.
The modeling of the two poverty rates in the Chilean dataset assumes that the spatial process follows a bivariate GMCAR model, from which Lin's coefficient is directly obtained based on a previous extension investigated in the literature. It is important to note that the GMCAR process is not the only method for modeling multivariate lattice data, as several other approaches have been explored in the literature. Our choice of the GMCAR model is motivated by its simplicity and the ease of interpreting the parameters in the covariance structures, both between and within the spatial components. For instance, the model allows for the incorporation of higher-order neighbors through a sequence of contiguity matrices.

The estimation of $\rho_{s,c}$ in \eqref{lattice_coef} was obtained using a Bayesian approach, which involved straightforward computation of the parameters required for the plug-in version presented in \eqref{bay_est}. It is worth noting that from a classical perspective, finding a statistical solution for this problem is not straightforward due to the absence of a well-defined asymptotic framework for lattice data.

From the application presented in Section \ref{sec:application} we remark that the proposed approach effectively captures the agreement between the two methods in the metropolitan areas. However, the discrepancies observed in Valparaíso warrant further investigation. We conjecture that the HT estimate is influenced by small counties with a limited number of samples  \citep{Casas:2016}.  Additionally, the sensitivity of the estimates to model or distribution misspecification remains an open problem that could be addressed in future research. Furthermore, the incorporation of more complex spatial weight structures and the investigation of the robustness of the coefficient deserve special attention in future studies.

{\footnotesize
  \noindent \textbf{Acknowledgements}~This work was partially supported by AC3E, UTFSM, under grant AFB240002. R. Vallejos also acknowledges financial support from CONICYT through the STIC-AMSUD program, grant 23STIC-02, and from Fondecyt, grant 1230012. The authors thank Felipe Osorio and Carolina Casas-Cordero for their insightful discussions that helped us better understand the CASEN survey.
 
  \noindent \textbf{Conflicts of interest}~The authors declare that they have no conflict of interest. \vspace{0.2cm}
  
  \noindent \textbf{Data availability}~The 2009 CASEN survey dataset used in this paper is available on the  the Ministerio de Desarrollo Social website
    \url{https://observatorio.ministeriodesarrollosocial.gob.cl/encuesta-casen-2009}.

\newpage

\appendix
\section{HT and SAE estimates}\label{sec: Appendix A}

The HT estimator is a method used to estimate both the total and the mean of a population within a stratified sample \citep{Maiti:2011}. Consider a finite population $U=\{1, \ldots, N\}$ within a specific county, where each individual $k\in U$ has a variable of interest $y_k$, $k \in U$ (such as income or an indicator of whether they fall below the poverty line). The goal is to analyze a function of $y_k$ denoted as $r_{\mathrm{HT}}=f(y_1, \ldots, y_N)$ which represents the county's poverty rate.

Consider a random sample $S$, selected from the population, taking values in  $s \subset U$ with probability $p(s)$. Additionally, let us denote the indicator function as
$$
I_k= \begin{cases}1, & k \in S, \\ 0, & k \notin S .\end{cases}
$$
Then, the inclusion probability for each individual $k$ can be written as
$$\pi_k=\mathbb{E}[I_k]=\sum_{k \in S} p(k).$$
The aim is to estimate both the total population of $y$, denoted by $Y=\sum_{k \in U} y_k$, and the population mean $\mu=\frac{1}{N}\sum_{k \in S} y_k $.

The Horvitz-Thompson estimator for the total population is 
$$
\widehat{Y}_{\mathrm{HT}}=\sum_{k \in S} \frac{y_k}{\pi_k},
$$
while the HT estimator for the population mean, which in the context of this paper represents the poverty rate, is given by
$$
r_{\mathrm{HT}}:=\widehat{\mu}_{\mathrm{HT}}=\frac{1}{N} \sum_{k \in S} \frac{y_k}{\pi_k}=\frac{\widehat{Y}_{\mathrm{HT}}}{N}.
$$
This process adjusts the observed values of individuals in the sample based on their probability of being included in the survey. By weighting each sampled individual inversely by their inclusion probability $\pi_k$, the HT estimator provides an unbiased estimate of both the total number of individuals living in poverty and the mean poverty rate for the entire population.

Small Area Estimation \citep{Rao:2015} refers to estimation methods used for populations where traditional approaches are inadequate due to high variability. In the context of poverty estimation, SAE improves the precision of estimates for each county by incorporating indirect information from external sources, thereby increasing accuracy. The methodology employed in this study for poverty estimation is based on the model developed by the U.S. Census Bureau, which is used to estimate poverty figures that serve as the basis for allocating public funds to localities.

The method consists of estimating the poverty rate at the county level, $r_{\mathrm{SAE}}$, as a weighted average between the direct poverty rate,  $r_{\mathrm{DIR}}$, and a synthetic poverty rate, $r_{\mathrm{SYN}}$, given by $$r_{\mathrm{SAE}}=(1-\lambda) r_{\mathrm{DIR}}+\lambda r_{\mathrm{SYN}},$$
where $$\lambda=\frac{\mathbb{V}\mathrm{ar}[r_{\mathrm{DIR}}]}{\mathbb{V}\mathrm{ar}[r_{\mathrm{DIR}}]+\mathbb{V}\mathrm{ar}[r_{\mathrm{SYN}}]}.$$

The direct poverty rate, $r_{\mathrm{DIR}}$, refers to the estimated incidence of poverty based on CASEN survey data. The synthetic poverty rate, $r_{\mathrm{SYN}}$, is derived from a linear prediction $r_{\mathrm{SYN}}=\boldsymbol{x}^{\top} \boldsymbol{\beta}$, where $\boldsymbol{x}$ represents auxiliary information from administrative records and census data for each county, and the coefficients, $\boldsymbol{\beta}$,  are estimated using a linear regression model.

\section{Additional figures and Tables}\label{sec: Appendix B}

\begin{table}[H]
    \centering
    \caption{Posterior mean of non-spatial GMCAR process parameters with $95\%$ HPD intervals.}
    \begin{tabular}{cccccc}
    \toprule 
     & $\widehat{\mu}_1$ & $\widehat{\mu}_2$ & $\widehat{\tau}_1$ & $\widehat{\tau}_2$ \\ 
     \midrule
     Santiago & $0.118$ & $0.120$ & $60.71$ & $59.76$ \\ \vspace{0.5cm}
     & $(0.082, 0.157)$ & $(0.067, 0.178)$ & $(30.65, 105.62)$ & $(29.95, 103.31)$ \\
     Valparaíso & $0.129$ & $0.123$ & $61.01$ & $51.43$ \\
     & $(0.085, 0.174)$ & $(0.031, 0.185)$ & $(28.51, 108.62)$ & $(22.75, 95.50)$ \\
    \bottomrule
    \end{tabular}
    \label{tab:non-spatial-summary}
\end{table}

\begin{table}[H]
    \centering
    \caption{Posterior mean of spatial GMCAR process parameters with $95\%$ HPD intervals.}
    \begin{tabular}{cccccc}
    \toprule 
     & $\widehat{\rho}_1$ & $\widehat{\rho}_2$ & $\widehat{\eta}_0$ & $\widehat{\eta}_1$ \\ 
     \midrule
     Santiago & $0.514$ & $0.444$ & $0.476$ & $0.152$ \\ \vspace{0.5cm}
     & $(0.035, 0.955)$ & $(0.023, 0.946)$ & $(-0.218, 1.198)$ & $(-0.121,0.454)$ \\
     Valparaíso & $0.376$ & $0.397$ & $0.091$ & $0.013$\\ 
     & $(0.017, 0.874)$ & $(0.019, 0.892)$ & $(-0.840, 1.037)$ & $(-0.488,0.515)$\\
    \bottomrule
    \end{tabular}
    \label{tab:spatial-summary}
\end{table}

\begin{figure}[H]
    \centering
    \includegraphics[width=0.8\linewidth]{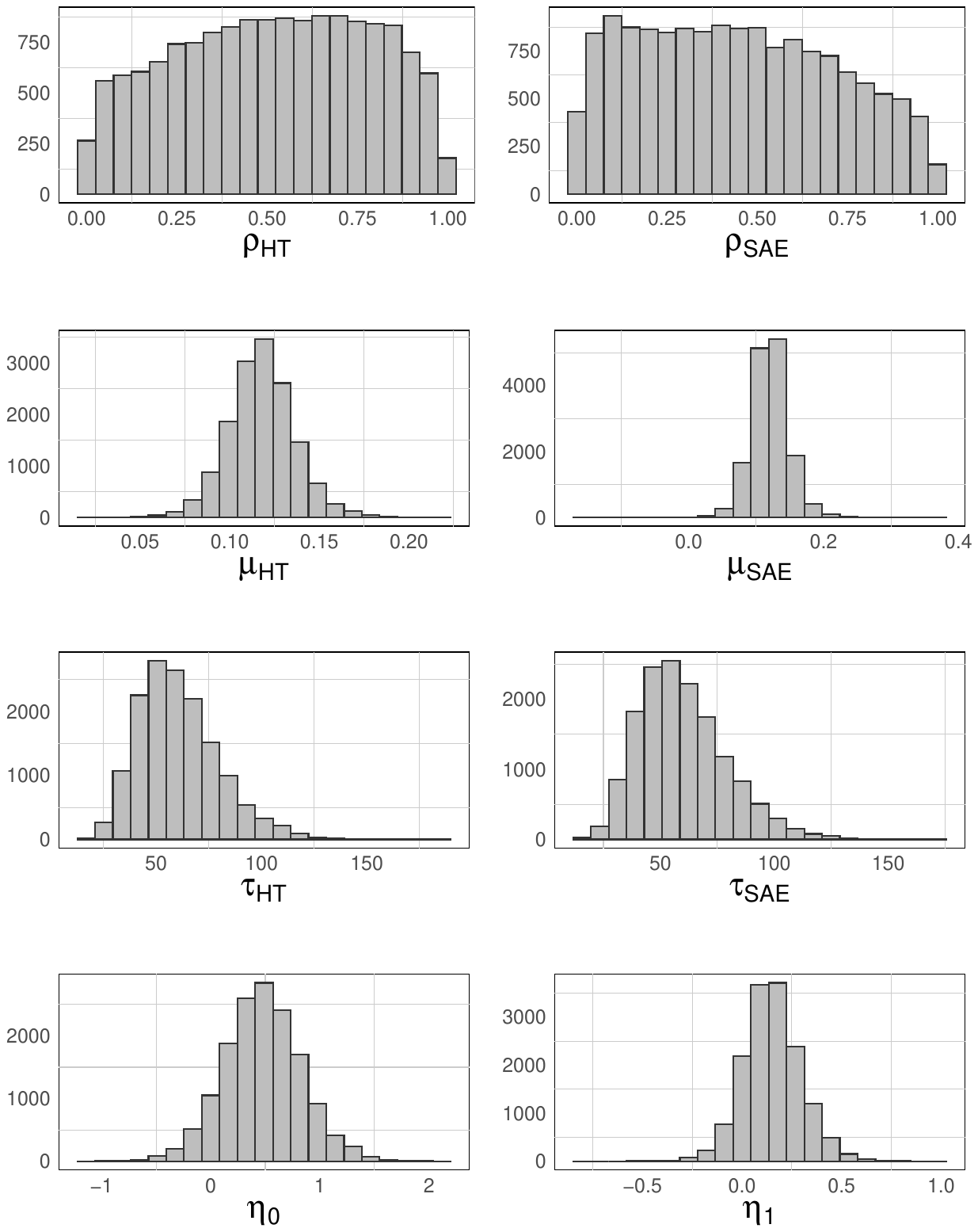}
    \caption{Posterior distribution of GMCAR model for Santiago Metropolitan Region.}
    \label{fig:add_figures_RM}
\end{figure}

\begin{figure}[H]
    \centering
    \includegraphics[width=0.8\linewidth]{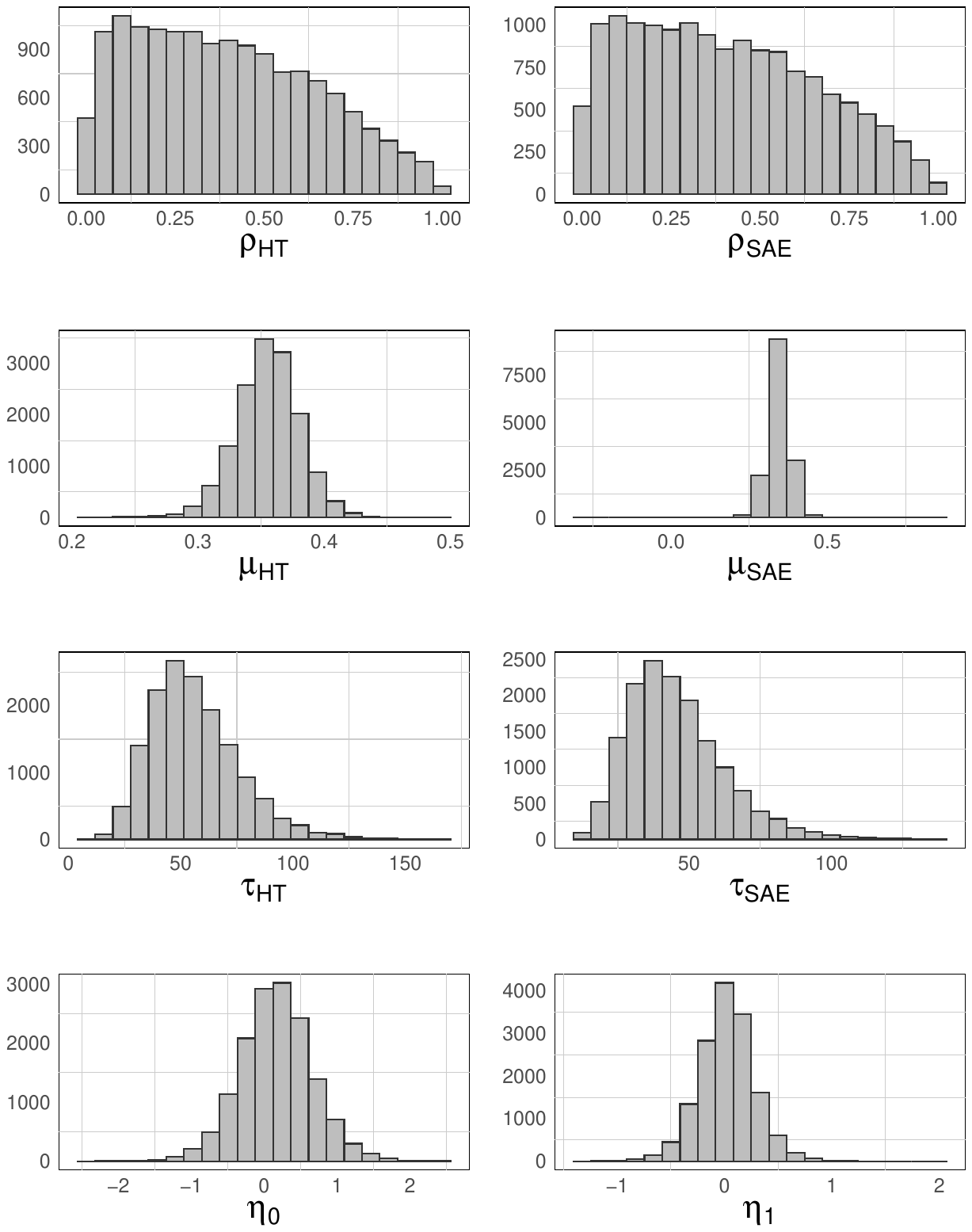}
    \caption{Posterior distribution of GMCAR model for Valparaiso Region.}
    \label{fig:add_figures_V}
\end{figure}

\end{document}